\documentclass[12pt]{article}

\usepackage{amsmath,mathrsfs,amsbsy,amsthm}
\usepackage{cite}
\usepackage{graphicx}
\usepackage{amssymb}
\usepackage{geometry}
\usepackage{amsfonts}
\usepackage{setspace}

\begin{document}
\doublespacing
	
\title{
	Phase-dependent kink collisions and dual critical-velocity branches
	in the complex sine-Gordon model
}

	\author{
		M. Mohammadi\thanks{Corresponding author: \texttt{physmohammadi@pgu.ac.ir}},
		F. Eizadbaksh, and V. Bagheri\\
		{\small Physics Department, Persian Gulf University, Bushehr 75169, Iran}\\
		{\small \texttt{farnazizadbakhsh@gmail.com}; \texttt{vbweb.company@gmail.com}}
	}
	\date{}
	\maketitle

\begin{abstract}
	
	The complex sine-Gordon (CSG) model contains an internal phase degree of freedom that strongly modifies the dynamics of its solitary-wave solutions. We present a numerical study of complex kink--kink collisions and determine how the final state depends jointly on the initial velocity and relative phase. In contrast with the elastic collisions of the real sine-Gordon model, the CSG system exhibits scattering, capture, long-lived bion formation, breather-like states, and emission of radiative profiles. The simulations reveal two distinct phase-dependent branches of critical velocity. In one branch, increasing the initial velocity promotes capture, whereas in the other it restores scattering. This dual structure highlights the rich velocity--phase dependence of the collision dynamics. We also compute the energy carried by radiative profiles and examine extreme values of the energy density, kinetic and gradient contributions, and field modulus at the collision center. These quantities show sharp transitions at critical points and provide sensitive diagnostics of phase-controlled dynamics. These results suggest that the relative phase behaves as an effective internal degree of freedom that plays an important role in the collision dynamics of complex solitons.
	
\end{abstract}


\section{Introduction}\label{sec1}

Nonlinear field theories supporting localized excitations constitute one of the most active areas of mathematical physics and nonlinear science. Solitary waves and topological defects arise in a broad range of physical systems, from particle physics and cosmology to condensed matter, nonlinear optics, and biological systems. Examples include domain walls, cosmic strings, monopoles, vortices, Skyrmions, and kink-like excitations, each providing important insights into nonperturbative phenomena and nonlinear dynamics \cite{cosmology1,cosmology2,Manton}.

Among one-dimensional relativistic models, the sine-Gordon (SG) theory occupies a distinguished position. Owing to its integrability, the SG equation possesses exact multi-soliton solutions and exhibits completely elastic collisions. Beyond its mathematical elegance, the SG model has found applications in diverse areas such as Josephson junctions, crystal dislocations, charge-density waves, nonlinear optics, and DNA dynamics \cite{condense2,Braun,optic3,bio}. Consequently, it has become a paradigmatic model for understanding nonlinear wave propagation and soliton interactions.

The situation becomes considerably richer in non-integrable scalar-field theories. Models such as the $\phi^4$, $\phi^6$, and double sine-Gordon systems exhibit intricate collision phenomena including critical velocities, capture processes, resonance windows, bion formation, and complex energy-exchange mechanisms \cite{phi44,phi46,phi47,phi61,DSG2}. These studies have demonstrated that even in single-component field theories the collision dynamics can be remarkably rich and sensitive to the initial conditions.

Considerable attention has been devoted to solitonic systems possessing internal degrees of freedom and nontrivial moduli spaces. In contrast to conventional real scalar-field models, where the motion of a solitary wave is essentially characterized by its position and velocity, multi-component and complex field theories may support additional collective coordinates associated with internal symmetries. Such internal moduli are known to generate phase-dependent interactions and novel scattering phenomena inaccessible to single-component systems. Examples include Q-kinks, Q-lumps, non-Abelian vortices, and orientational solitons carrying internal modes \cite{Abraham,Leese,Eto,CTOPO,Nitta2022,tONG}. Understanding the role of these internal degrees of freedom in nonlinear wave interactions remains an important problem in classical field theory.

The complex sine-Gordon (CSG) model provides a particularly attractive framework for studying these issues. As a complex extension of the conventional sine-Gordon theory, the CSG system possesses a global $U(1)$ symmetry and admits solitary-wave solutions endowed with an internal phase degree of freedom. In addition to complex topological kinks, the model supports non-topological configurations and radiative excitations \cite{MR}. The existence of an internal phase considerably enlarges the dynamical landscape and raises the possibility of phase-controlled interactions, energy redistribution among field components, and collision outcomes absent in real scalar-field theories.

Previous investigations of the CSG model \cite{MR} have mainly focused on exact solutions and general properties of the theory. However, a systematic study of the combined influence of the relative phase and collision velocity on complex kink interactions remains largely unexplored. Since the relative phase provides an additional internal degree of freedom, one naturally expects the collision dynamics to exhibit a genuinely two-parameter character, depending not only on the initial velocity but also on the phase difference between the colliding kinks.

The purpose of the present work is to investigate this velocity–phase dependence through extensive numerical simulations of kink--kink collisions in the CSG model. By systematically varying both the initial velocity and the relative phase, we demonstrate that the collision outcomes exhibit strong phase sensitivity and reveal phenomena that have no counterparts in conventional real scalar-field theories. In particular, we identify two distinct phase-dependent branches of critical velocities separating scattering and capture regimes, investigate phase-controlled collision outcomes involving breather and bion states within the capture region, and analyze the energy carried by the emitted radiative profiles. Furthermore, we show that several extreme quantities at the collision center provide sensitive indicators of phase-dependent critical behavior.

These results indicate that the relative phase plays an important role in governing the collision dynamics and substantially enriches the phenomenology of complex kink interactions. In this sense, the complex sine-Gordon model provides a simple framework for studying phase-dependent soliton interactions in field theories with internal degrees of freedom.

The paper is organized as follows. In Sec.~2, we briefly review the real and complex sine-Gordon models and discuss their conserved quantities. Section~3 introduces the different classes of solitary-wave solutions. The numerical scheme and initial conditions are described in Sec.~4. The main numerical results are presented in Sec.~5, while Sec.~6 is devoted to the analysis of several extreme quantities at the collision center. Finally, Sec.~7 summarizes the main conclusions and outlines possible directions for future investigations.

\section{Real and Complex Sine-Gordon Models}\label{sec2}

	The Lagrangian density of the nonlinear, relativistic real sine-Gordon (SG) system in \(1+1\)-dimensional space-time is given by
	\begin{equation} \label{lag}
		\mathcal{L}_{r} = \frac{1}{2}\partial_\mu \varphi \partial^\mu \varphi -  \left(1 - \cos(\varphi)\right),
	\end{equation}
	where \(\varphi\) is a real scalar field and \(\partial_\mu\) (\(\partial^\mu\)) denotes the covariant (contravariant) relativistic derivative. The corresponding field equation is
	\begin{equation} \label{de}
		\partial_\mu \partial^\mu \varphi = \ddot{\varphi} - \varphi'' = -\sin(\varphi),
	\end{equation}
	where the dot and prime denote partial derivatives with respect to time and space, respectively. Throughout this work, we adopt natural units and set the speed of light \(c = 1\) for simplicity.
	
	The energy--momentum tensor corresponding to the Lagrangian density \eqref{lag}, derived from Noether's theorem, is
	\begin{equation} \label{rt}
		T^{\mu \nu} = \partial^{\mu} \varphi \, \partial^{\nu} \varphi - g^{\mu \nu} \mathcal{L},
	\end{equation}
	where \(g^{\mu \nu}\) is the Minkowski space-time metric with signature \((+,-)\), i.e., \(g^{00} = -g^{11} = 1\), \(g^{01} = g^{10} = 0\). The \(T^{00}\) component represents the energy density:
	\begin{equation} \label{red}
		\varepsilon(x,t) = \frac{1}{2}(\dot{\varphi}^2 + \varphi'^2) + 1 - \cos(\varphi).
	\end{equation}
	
	The real SG system \eqref{lag} admits well-known solitary-wave solutions in the form of kinks (positive sign) and antikinks (negative sign),
	\begin{equation} \label{kin}
		\varphi_{v} = 4 \arctan\left(e^{\pm \gamma(x - vt - a)}\right) + 2N\pi,
	\end{equation}
	where \(N\in\mathbb{Z}\), \(v\) is the velocity of the kink (antikink), \(a\) is the initial position, and \(\gamma = 1/\sqrt{1 - v^2}\) is the Lorentz factor. These solutions exhibit classical soliton properties, such as stability and elastic collisions.
	
	In contrast, the Lagrangian density of the complex sine-Gordon (CSG) system generalizes \eqref{lag} by replacing the real scalar field $\varphi$ with a complex scalar field \(\phi\):
	\begin{equation} \label{lag2}
		\mathcal{L}_{c} = \frac{1}{2}\partial_\mu \phi \, \partial^\mu \phi^{*} -  \left(1 - \cos(|\phi|)\right).
	\end{equation}
	The corresponding field equation, energy--momentum tensor, and energy density take the forms
	\begin{align}
		\label{col0}
		\partial_\mu \partial^\mu \phi &= \ddot{\phi} - \phi'' = -\frac{\phi}{|\phi|} \sin(|\phi|),\\
		\label{col2}
		T^{\mu \nu} &= \partial^{\mu} \phi^* \partial^{\nu} \phi - g^{\mu \nu} \mathcal{L}_{c},\\
		\label{col3}
		T^{00} &= \frac{1}{2}|\dot{\phi}|^2 + \frac{1}{2}|\phi'|^2 + 1 - \cos(|\phi|) = k(x,t) + u(x,t) + p(x,t),
	\end{align}
	where the kinetic, gradient, and potential energy densities are defined as \(k(x,t) = \frac{1}{2}|\dot{\phi}|^2\), \(u(x,t) = \frac{1}{2}|\phi'|^2\), and \(p(x,t) = 1 - \cos(|\phi|)\), respectively.
	
	Due to the global \(U(1)\) symmetry of the Lagrangian \eqref{lag2}, the system admits a conserved Noether current \(j^\mu\) and associated charge \(q\):
	\begin{align}
		\label{cv1}
		j^\mu &= i\eta \left(\phi^* \partial^\mu \phi - \phi \partial^\mu \phi^*\right),\\
		\label{cv2}
		q &= \int_{-\infty}^{+\infty} j^0 \, dx = \int_{-\infty}^{+\infty} i\eta \left(\phi^* \dot{\phi} - \phi \dot{\phi}^*\right) dx,
	\end{align}
	where \(\partial_\mu j^\mu = 0\), and \(\eta\) is a real normalization constant.
	In addition, the system supports a conserved topological current \(J^\mu\) and charge \(Q\), applicable to both the real and complex SG models:
	\begin{align}
		\label{bv1}
		J^\mu &= C \epsilon^{\mu \nu} \partial_\nu \phi,\\
		\label{bv2}
		Q &= \int_{-\infty}^{+\infty} J^0 \, dx = C[\phi(+\infty) - \phi(-\infty)],
	\end{align}
	where \(\epsilon^{\mu \nu}\) is the antisymmetric Levi-Civita tensor, \(\partial_\mu J^\mu = 0\), and \(C\) is a positive normalization constant, chosen here as \(C = 1/2\pi\).
	
	The existence of a global $U(1)$ symmetry distinguishes the CSG model from its real counterpart and introduces an additional internal degree of freedom associated with the phase of the complex field. Although this phase does not affect the static energy of isolated kinks, it plays a crucial role in their interactions and gives rise to considerably richer collision dynamics. In this sense, the CSG model provides one of the simplest field-theoretical settings in which the influence of internal moduli on soliton interactions can be investigated.

	\section{Solitary Wave Solutions} \label{sec3}
	
	The complex sine-Gordon (CSG) system admits three main classes of solitary wave solutions: complex kinks, radiative profiles, and Q-balls.
	
	\subsection{Complex Kink Solutions} \label{mn}
	
	Inspired by the soliton solutions of the real SG system \eqref{kin}, the complex kink solutions of the CSG system \eqref{lag2} can be introduced as:
	\begin{equation} \label{ckin}
		\phi_{v}(x,t) = |\varphi_{v}| e^{i\theta} = \left|4\arctan(e^{+\gamma(x - vt - a)}) + 2N\pi\right| e^{i\theta},
	\end{equation}
	where \(\theta\) is an arbitrary constant phase. We note that the antikink counterpart is not separately considered here, and only the positive sign in Eq.~\eqref{kin} (kink) is employed.
	
	More precisely, the complex field \(\phi\) consists of real and imaginary components, \(\phi = \phi_r + i\phi_i\), each of which may individually possess kink or antikink structure with distinct topological charges:
	\begin{equation}\label{cvf}
		\phi \equiv
		\begin{cases}
			\phi_r = |\varphi_{v}| \cos(\theta),\quad Q_r = \cos(\theta),\\
			\phi_i  = |\varphi_{v}| \sin(\theta),\quad Q_i = \sin(\theta).
		\end{cases}
	\end{equation}
	If \(Q_r > 0\) (\(Q_r < 0\)), the real component \(\phi_r\) exhibits a sub-kink (sub-antikink) profile. Similarly, for \(Q_i > 0\) (\(Q_i < 0\)), the imaginary component \(\phi_i\) forms a sub-kink (sub-antikink). Thus, Eq.~\eqref{ckin} describes a continuous family of degenerate states parameterized by \(\theta\).
	
	Pure kink or antikink configurations (with either \(\phi_r = 0\) or \(\phi_i = 0\)) arise only for special phase values. For example, \(\theta = 0\) or \(\pi/2\) yields a pure kink, while \(\theta = \pi\) or \(3\pi/2\) gives a pure antikink.
	In general, for an arbitrary phase \(\theta\), a complex kink solution comprises two subfields, \(\phi_r\) and \(\phi_i\), which together form nontrivial configurations: a sub-kink in one component and a sub-antikink in the other, sub-kinks in both components, or sub-antikinks in both. This implies that the kink--antikink duality of the real case does not directly carry over to the complex version. This richer internal structure highlights the enhanced symmetry of the CSG system.

	For the complex kink solutions \eqref{ckin}, the phase \(\theta\) is spatially and temporally constant, and hence the components of the energy--momentum tensor \eqref{col3} are independent of \(\theta\). Therefore, both the total energy and momentum of the kink, obtained by integrating \(T^{00}\) and \(T^{01}\), are independent of the phase. For a moving complex kink, the standard relativistic energy--momentum relations hold:
	\begin{align}
		\label{fv}
		E_v &= \gamma E_0 = \int_{-\infty}^{+\infty} T^{00} \, dx, \\
		\label{fv2}
		P &= \gamma m_0 v = \int_{-\infty}^{+\infty} T^{10} \, dx,
	\end{align}
	where \(E_0 = m_0 c^2\) is the rest energy, and \(m_0\) is the rest mass of the kink.
	
	\subsection{Radiative Profiles} \label{mn2}

	For any spatial and temporal configuration of the phase field, provided that the complex scalar field takes the following form:
	\begin{equation} \label{rjk}
		\phi = (n\pi) e^{i\theta(x,t)}, \quad n = 1, 2, 3, \ldots,
	\end{equation}
	or, equivalently, when the real and imaginary parts of the complex field \(\phi\) satisfy:
	\begin{equation} \label{vv}
		|\phi| = \sqrt{\phi_r^2 + \phi_i^2} = n\pi, \quad n = 1, 2, 3, \ldots,
	\end{equation}
	the right-hand side of the field equation \eqref{col0} vanishes, and the resulting equation becomes a homogeneous linear wave equation with general traveling-wave solutions:
	\begin{equation} \label{cv}
		\phi(x,t) = \phi(x \pm t).
	\end{equation}
	These solutions are referred to as ``\textit{radiative profiles}''.

	It can be shown that radiative profiles satisfy the standard energy--momentum relation for massless entities (e.g., photons):
	\begin{equation} \label{af}
		E = Pc,
	\end{equation}
	where
	\begin{equation} \label{ad}
		E = \int_{-\infty}^{+\infty} T^{00} \, dx, \quad P = \int_{-\infty}^{+\infty} T^{10} \, dx.
	\end{equation}
	Radiative profiles can be either topological or non-topological. As an example of a non-topological radiative solution, we introduce:
	\begin{equation} \label{sc}
		\phi_r = e^{-(x - t)^2}, \quad \phi_i = \sqrt{4\pi^2 - e^{-2(x - t)^2}}.
	\end{equation}
	An example of a topological radiative profile is given by:
	\begin{equation} \label{sd}
		\phi_r = \sqrt{4\pi^2 - \tanh^2(x + t)}, \quad \phi_i = \tanh(x + t).
	\end{equation}

	\subsection{Q-ball Solutions} \label{mn3}

	The complex sine-Gordon system also admits non-topological localized solutions known as Q-balls \cite{Tsumagari,Bowcock,MohammadiQ}. These solutions correspond to time-periodic, spatially localized field configurations carrying a conserved Noether charge.
	Although Q-balls are an important class of solutions in complex scalar field theories, they are not relevant to the present study, which focuses exclusively on topological complex kink configurations and their collision dynamics. Therefore, Q-ball solutions are not investigated further in this work.

	\section{Initial Conditions and Numerical Method} \label{sec4}

	\begin{figure}[htp]
		\centering
		\includegraphics[width=120mm]{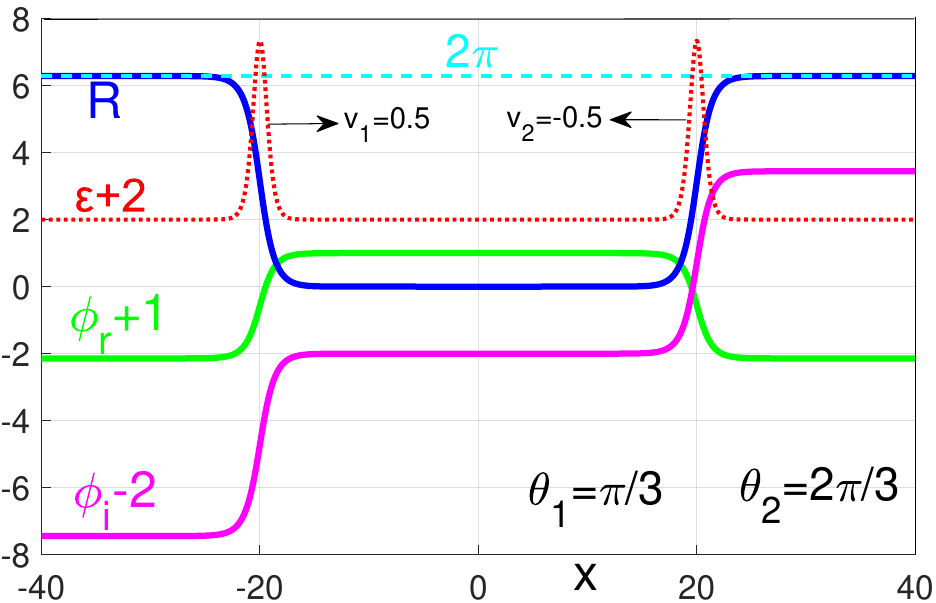}
		\caption{The real and imaginary components, energy density, and modulus of a pair of complex kinks, plotted using Eq.~\eqref{jg} with parameters \(v_1 = -v_2 = 0.5\), \(b = -a = 20\), \(\theta_1 = \pi/3\), and \(\theta_2 = 2\pi/3\). For clarity, each curve has been vertically offset by an arbitrary constant to prevent overlap.
		}
		\label{mj}
	\end{figure}

	The main objective of this study is to determine how collisions between two complex kinks depend on the initial relative phase and velocity.
	Due to the nonlinearity of the dynamical equation~\eqref{col0}, a simple linear superposition of two complex kink solutions is not an exact solution. However, if the initial separation is sufficiently large, such a linear combination, augmented by a constant shift, provides an accurate approximate initial condition. For kinks initially centered at \(a\) and \(b\), with velocities \(v_1\) and \(v_2\), we use
	\begin{equation} \label{jg}
		\phi = 4 \arctan\big(e^{\gamma_1(x - v_1 t - a)}\big)e^{i\theta_1} + 4 \arctan\big(e^{\gamma_2(x - v_2 t - b)}\big)e^{i\theta_2} - 2\pi e^{i\theta_1}.
	\end{equation}
	Here, the constant term \( -2\pi e^{i\theta_1} \) ensures that the modulus field \( R = |\phi| \) interpolates between the adjacent vacua \(0\) and \(2\pi\). This adjustment localizes the energy density into two distinct peaks, representing two solitonic structures approaching one another (see Fig.~\ref{mj}).

	For numerical simulations, we employ the second-order central finite-difference form of the complex wave equation~\eqref{col0},
	\begin{equation} \label{ph}
		\frac{1}{k^2}(\phi_{n,m+1} + \phi_{n,m-1} - 2\phi_{n,m}) - \frac{1}{h^2}(\phi_{n+1,m} + \phi_{n-1,m} - 2\phi_{n,m}) = -\frac{\phi_{n,m}}{|\phi_{n,m}|} \sin(|\phi_{n,m}|),
	\end{equation}
	where \(h\) and \(k\) are the spatial and temporal discretization steps, respectively. In all simulations we set \(h = 0.02\), \(k = 0.019\), and \(b = -a = 20\). To suppress spurious reflections, the spatial domain is chosen wide enough that the boundaries remain beyond the furthest distance traveled by the wavefronts during the simulated time interval.

	\section{Numerical Results} \label{sec5}
	
	We studied the collisions between pairs of complex kinks for various phase differences \(\Delta\theta = \theta_2 - \theta_1\) and initial velocities.
	As expected from the global \(U(1)\) symmetry, and confirmed by our numerical simulations, the collision outcomes depend on the phase difference between the kinks rather than on the two absolute phases separately.
	Therefore, without loss of generality, we fix \(\theta_1 = 0\) and denote the phase difference simply as \(\theta = \theta_2\)\footnote{Although ``phase difference'' is the precise term, we use ``phase'' below for brevity.}.
	Our numerical results further reveal that the collision outcomes are symmetric with respect to \(\theta\) and \(-\theta\). Consequently, the numerical investigation is restricted to the interval \(0 < \theta < \pi\).
	A key immediate observation is that in collisions of out-of-phase complex kink pairs, i.e., for \(\theta \neq 0\) and \(\theta \neq \pi\), at least two radiative profiles are emitted from the collision site and travel in opposite directions at the speed of light.
	In contrast, for in-phase complex kink pairs (\(\theta = 0\) or \(\theta = \pi\)), the collision outcomes coincide exactly with those of kink-kink or kink-antikink collisions in the real SG system.
	
	\subsection{Critical Velocity}
	
	\begin{figure}[htp]
		\centering
		\includegraphics[width=120mm]{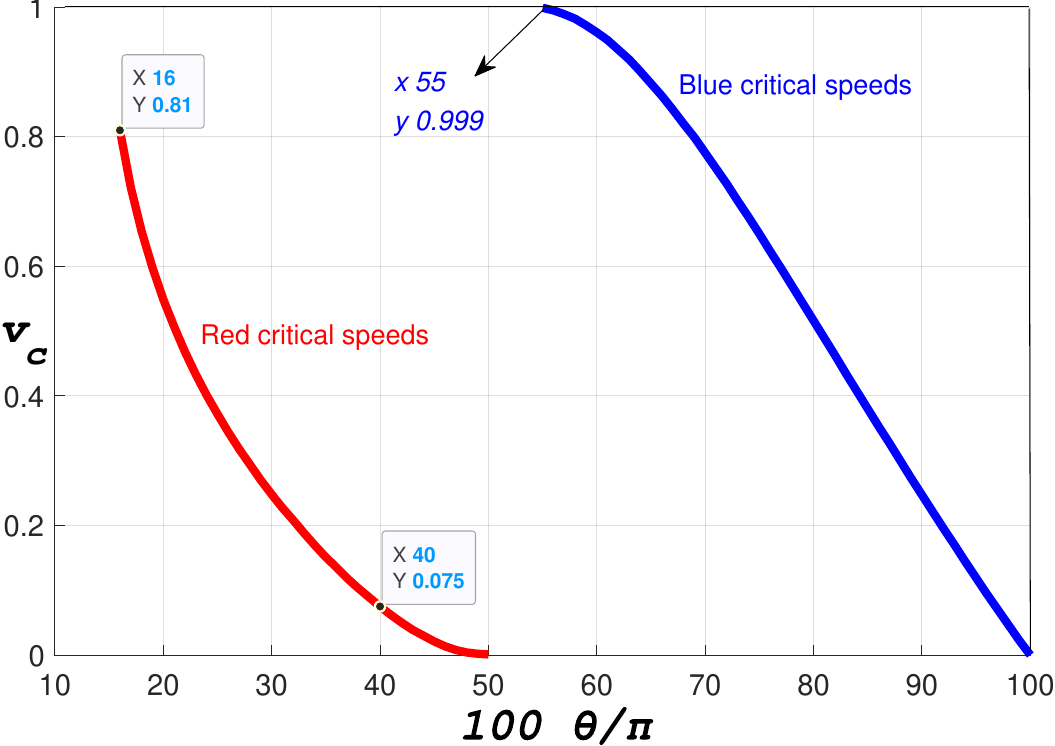}	
		\caption{Critical speed \(v_c\) versus phase \(\theta\). Numerical data were obtained for two distinct intervals: \(0.16\pi < \theta < 0.5\pi\) and \(0.55\pi < \theta < \pi\).}
		\label{bn}
	\end{figure}
	
	In systems admitting kink solutions, such as the \(\varphi^4\) model, the \textit{critical velocity} \(v_c\) is usually defined as the threshold initial velocity above which kink--antikink collisions result in scattering rather than capture.
	For the integrable real SG model, this concept is not applicable.
	However, in the CSG system, the notion of critical speed manifests in a nontrivial manner.
	Our numerical simulations indicate that \(v_c\) can be well-defined for certain phase intervals, while for others, it approaches either zero or the speed of light, hindering precise numerical identification.
	Figure~\ref{bn} presents the critical velocity as a function of \(\theta\), divided into two distinct regimes indicated by red and blue colors.
	This color distinction corresponds to two different dynamical meanings of \(v_c\) across phase ranges.
	
\begin{figure}[htp]
		\centering
		\begin{tabular}{cc}
			\includegraphics[width=70mm]{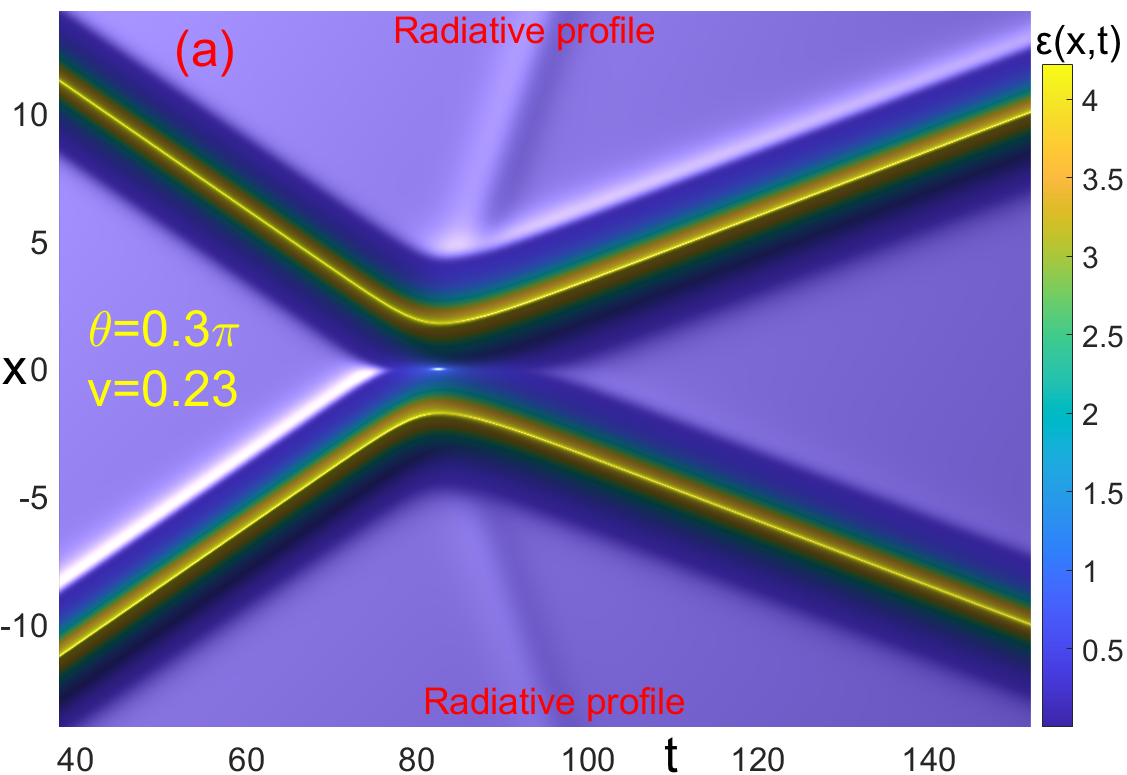} &
			\includegraphics[width=70mm]{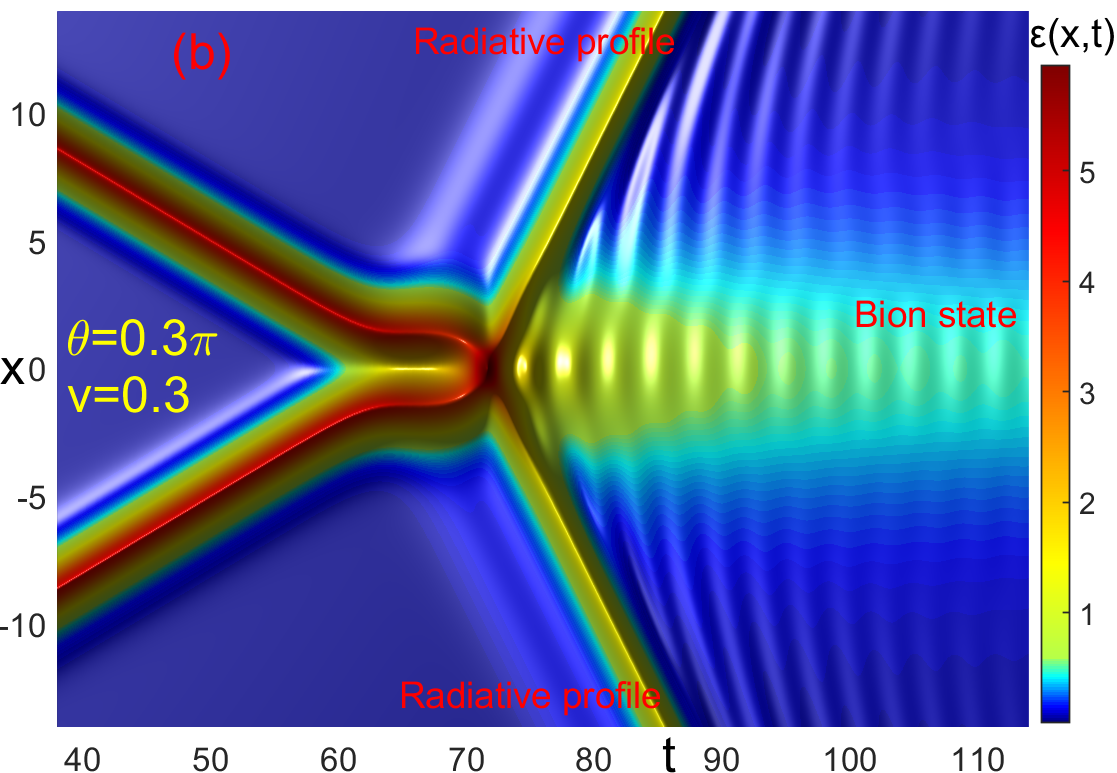} \\
			\includegraphics[width=70mm]{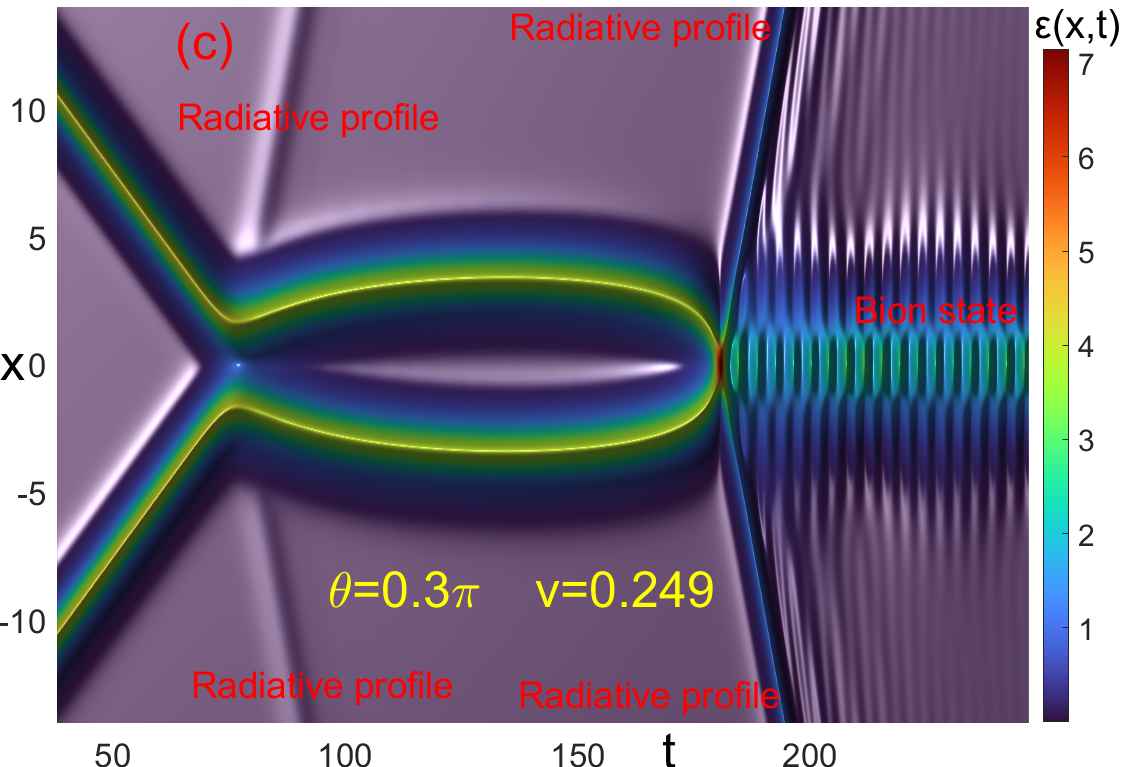} &
			\includegraphics[width=70mm]{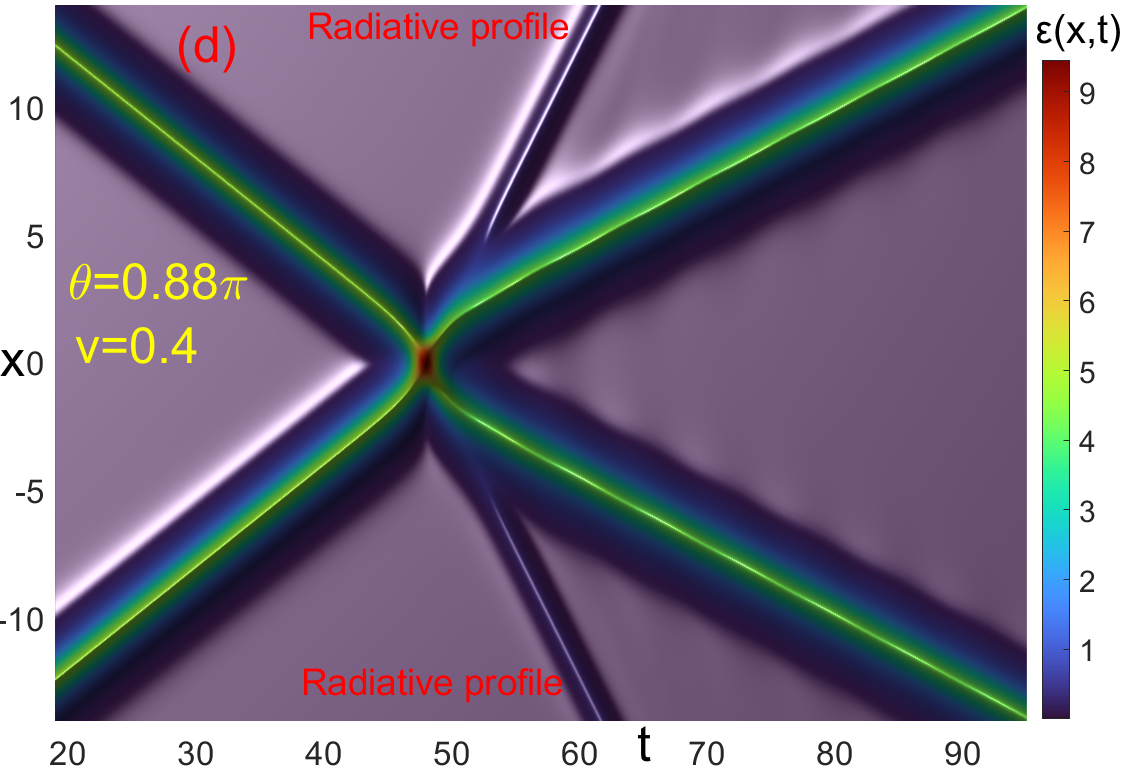} \\
			\includegraphics[width=70mm]{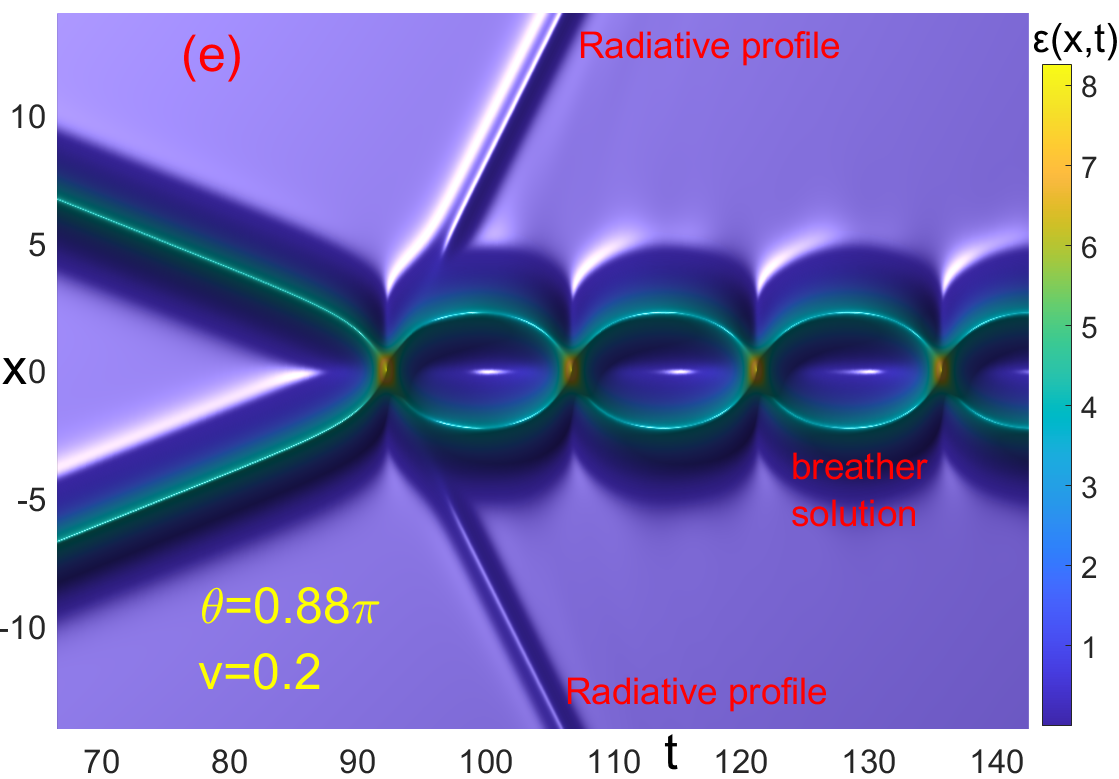} &
			\includegraphics[width=70mm]{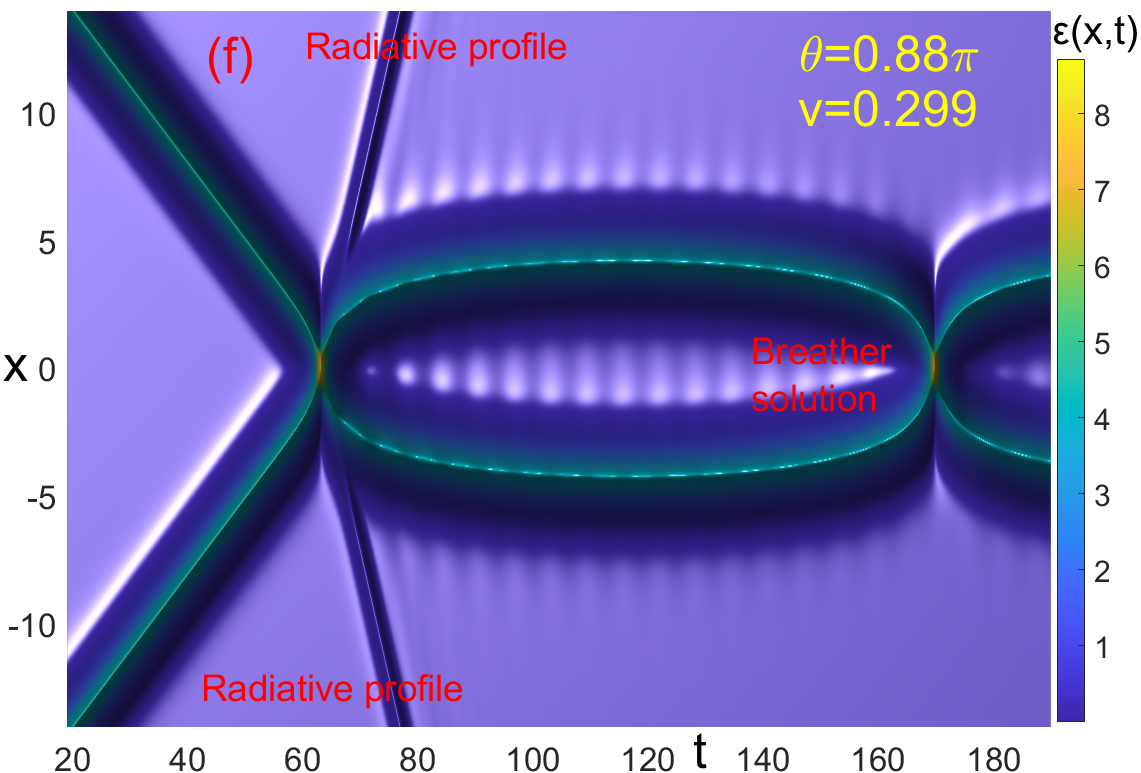}
		\end{tabular}
	\caption{Energy density profiles in six collision scenarios, grouped by phase. Each phase includes collisions at velocities below, above, and near the corresponding critical velocity.}
		\label{hjn}
	\end{figure}

	Remarkably, the red region corresponds to phases where high-energy collisions lead to capture, while low-energy collisions result in scattering.
	Here, the critical velocity is interpreted as the velocity above which capture becomes the only outcome, in contrast to the conventional interpretation where velocities above \(v_c\) correspond to scattering.
	Accordingly, we classify the critical velocities into two categories:
	in the red region, initial velocities exceeding \(v_c\) lead to capture;
	in the blue region, initial velocities exceeding \(v_c\) lead to scattering.
	For instance, the critical velocity at \(\theta=0.3\pi\) is approximately \(v=0.249\) and belongs to the red category, whereas at \(\theta=0.88\pi\) it is approximately \(v=0.299\) and belongs to the blue category (see Fig.~\ref{hjn}).

	For the phase ranges \(0.16\pi < \theta < 0.5\pi\) and \(0.55\pi < \theta < \pi\), the critical velocities correspond to the red and blue categories, respectively.
	In the ranges \(0 < \theta < 0.16\pi\) and \(0.5\pi < \theta < 0.55\pi\), numerical determination of \(v_c\) is hindered by computational constraints, with the critical velocity tending close to 1 or 0 where applicable.
	
	\subsection{Bions and Breathers}
	
	\begin{figure}[htp]
		\centering
		\includegraphics[width=120mm]{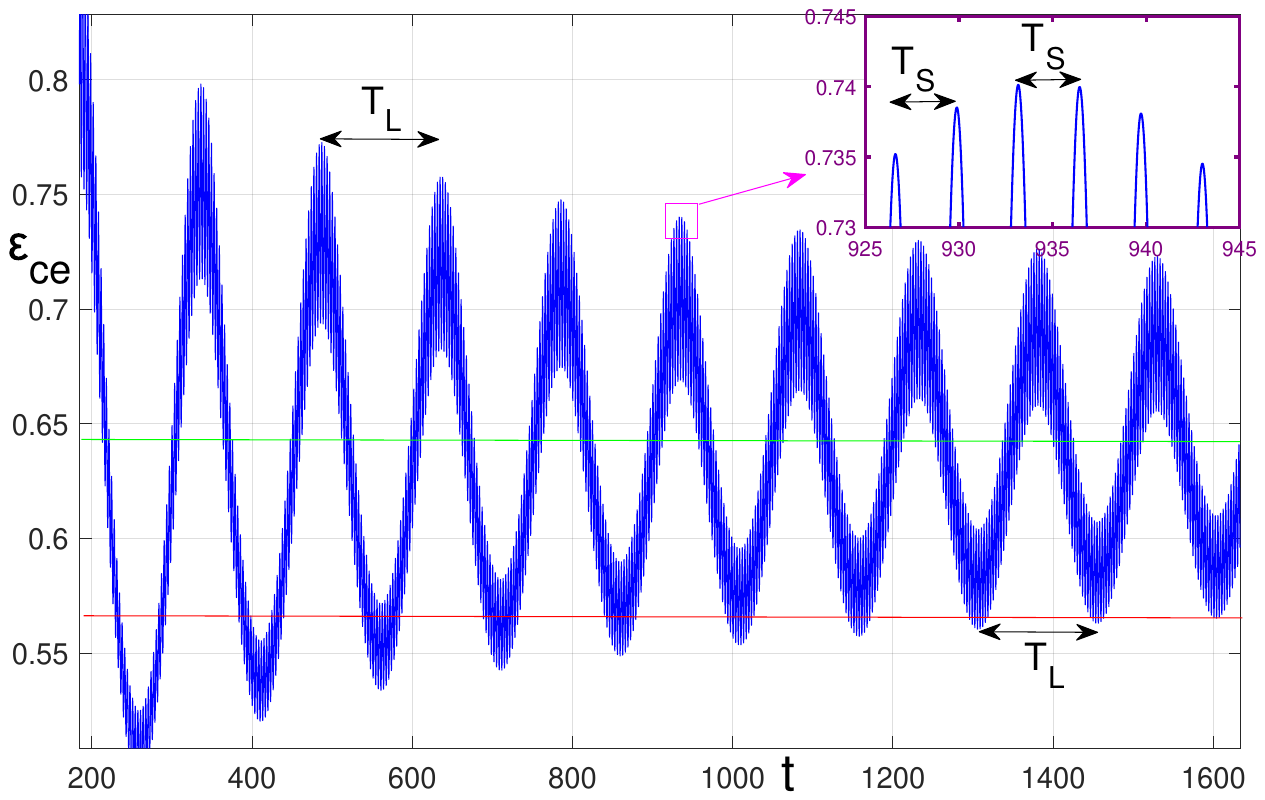}
		\caption{Central energy density \(\varepsilon_{ce}\) of a bion formed by the collision of two complex kinks with parameters \(\theta = 0.2\pi\) and \(v = 0.6\). Two oscillation modes with distinct periods, short and long, are identified.}
		\label{nb}
	\end{figure}
	
	\begin{figure}[htp]
		\centering
		\includegraphics[width=120mm]{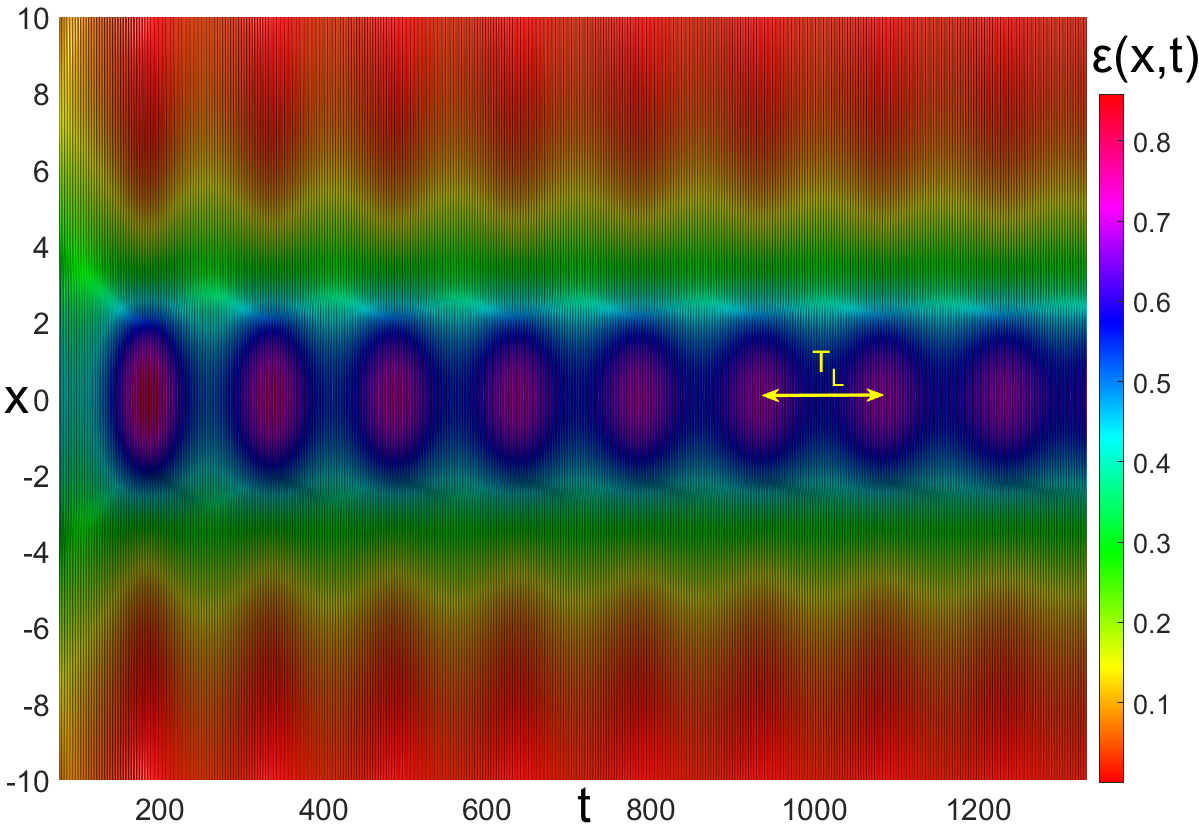}
		\caption{Three-dimensional energy density evolution during the collision at \(\theta=0.2\pi\) and \(v=0.6\). Narrow bands illustrate short-period oscillations within the bion region.}
		\label{vb}
	\end{figure}
	
	In out-of-phase collisions, if the kinks capture each other, the resultant localized oscillatory structure can be either a \textit{bion} or a \textit{breather-like} state.
	While breather-like states preserve a regular oscillatory profile throughout the simulated time window (see Figs.~\ref{hjn}(e) and \ref{hjn}(f)), bions represent unstable energy accumulations that gradually radiate energy away (see Figs.~\ref{hjn}(b) and \ref{hjn}(c)).
	Depending on the phase and initial velocity, bions may either radiate nearly all their energy and vanish (pair annihilation) or relax toward long-lived breather-like states after losing part of their energy.
	Typically, annihilation occurs at relatively high collision velocities when capture is present.

	The energy of bions is primarily concentrated at the collision site but is spatially more extended compared to breathers.
	Although the energy slowly decays through radiation from the outer regions, an ongoing interaction with internal oscillations leads to energy exchange between the central peak and the surrounding broadened region.
	Two oscillation modes are generally present: one with a long period \(T_L\) and another with a short period \(T_S\), as demonstrated in Figs.~\ref{nb} and \ref{vb}.
	For example, a collision at \(\theta=0.2\pi\) and \(v=0.6\) leads to a bion with \(T_L \approx 149.3\) and \(T_S \approx 3.28\).
	
	Tracking the temporal evolution of the energy density at the collision center, \(\varepsilon_{ce} = \varepsilon(0,t)\), provides an effective criterion for determining whether a bion eventually stabilizes into a breather-like state or disappears.
	For the case shown in Fig.~\ref{nb}, the oscillatory envelope narrows around a nonzero mean (green line), indicating stabilization rather than disappearance. Equivalently, the central energy density never approaches zero.
	Conversely, if \(\varepsilon_{ce}\) tends toward zero after the collision, the bion has dissipated its energy and the outcome may be interpreted as annihilation.
	A practical reason for using \(\varepsilon_{ce}\) instead of the total energy to characterize the bion fate is the difficulty of defining a sharp spatial boundary for an extended oscillatory object.

\begin{figure}[htp]
		\centering
		\begin{tabular}{cc}
			\includegraphics[width=51mm]{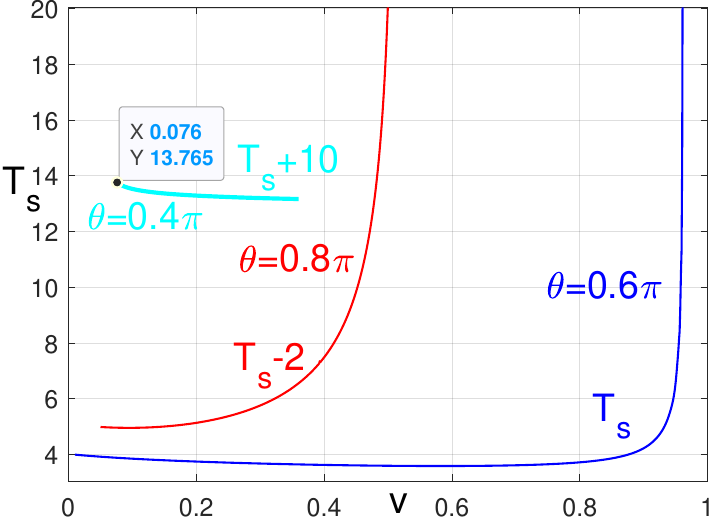} &
			\includegraphics[width=52mm]{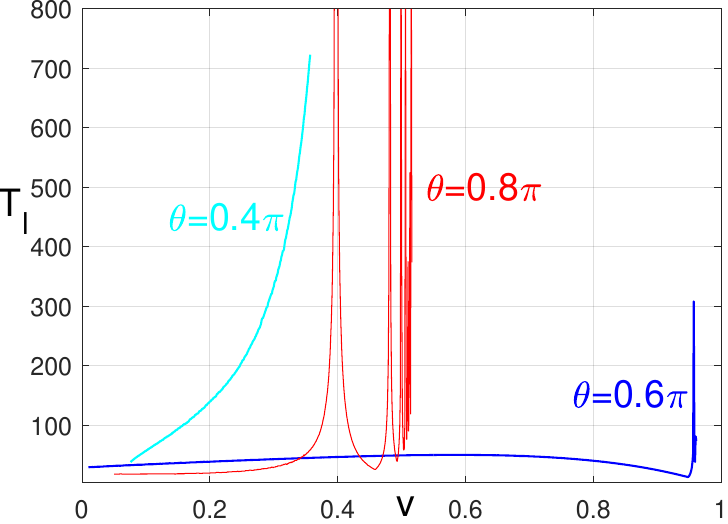} \\
			\includegraphics[width=50mm]{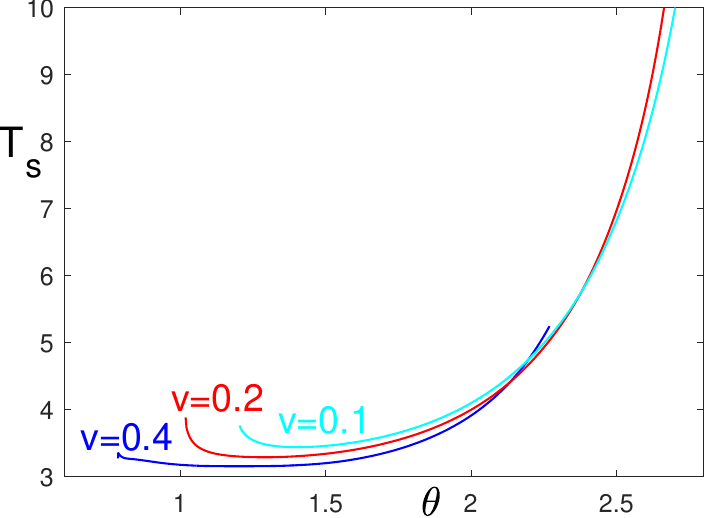} &
			\includegraphics[width=52mm]{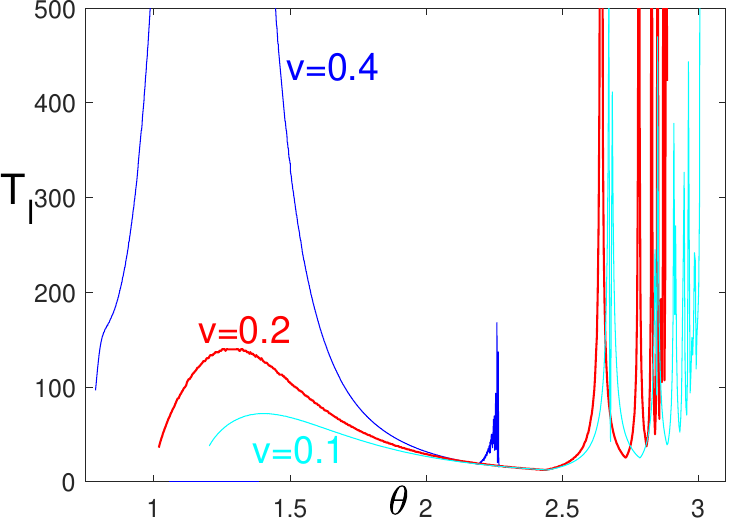}
		\end{tabular}
		\caption{Variations of the oscillation periods \(T_L\) and \(T_S\) as functions of velocity at three fixed phases (top row) and as functions of phase at three fixed velocities (bottom row).}
		\label{dop}
	\end{figure}

	Both \(T_L\) and \(T_S\) depend on phase and velocity.
	While \(T_S\) is numerically straightforward to extract, \(T_L\) requires long-time simulation to accurately resolve.
	Figure~\ref{dop} shows the dependence of these periods on velocity at three fixed phases and on phase at three fixed velocities, focusing on the capture regions identified in Fig.~\ref{bn}.

	\subsection{Radiative Profile Energy}

	\begin{figure}[htp]
		\centering
		\begin{tabular}{c}
			\includegraphics[width=125mm]{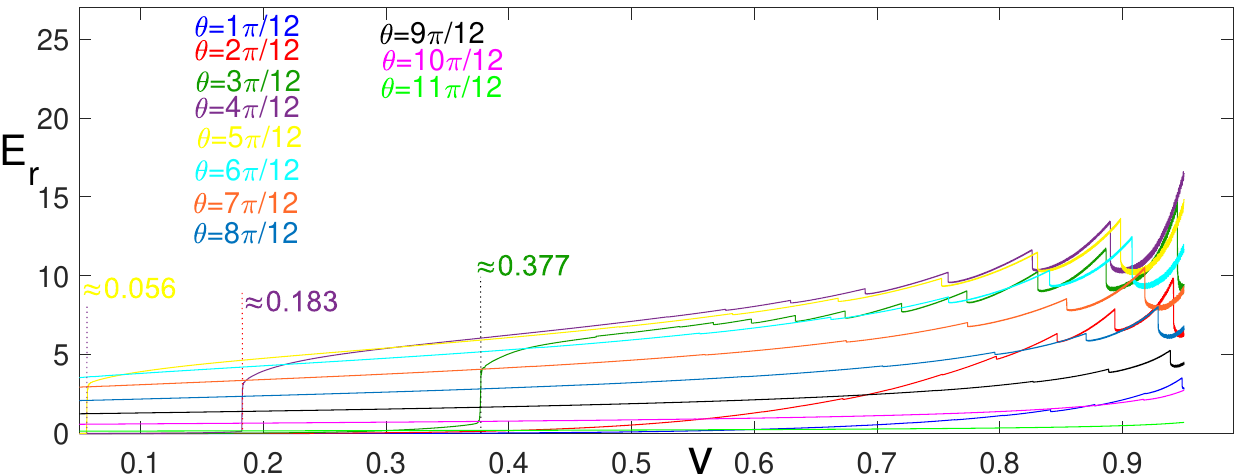}\\
			\includegraphics[width=125mm]{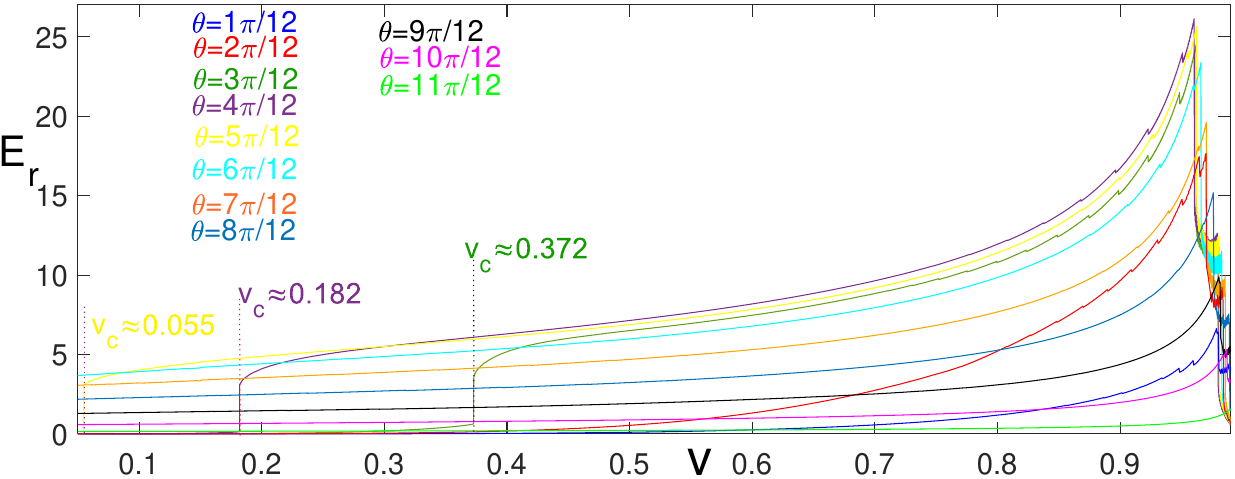}
		\end{tabular}
		\caption{Radiative profile energy emitted from collisions as a function of velocity for selected fixed phases. The top panel shows the energy measured at $100$ time units post-collision, while the bottom panel corresponds to measurements at $220$ time units post-collision.}
		\label{ft}
	\end{figure}
	
	As discussed earlier, in out-of-phase collisions, whether they result in capture or scattering, at least one pair of radiative profiles is emitted from the collision center and travels in opposite directions.
	The shape and energy of these emitted profiles depend significantly on both the initial velocity and the relative phase of the collision.
	To provide a more accurate understanding of the dependence of radiative profile energy on these parameters, the graphs in Fig.~\ref{ft} have been generated numerically.

\begin{figure}[htp]
		\centering
		\begin{tabular}{c}
			\includegraphics[width=120mm]{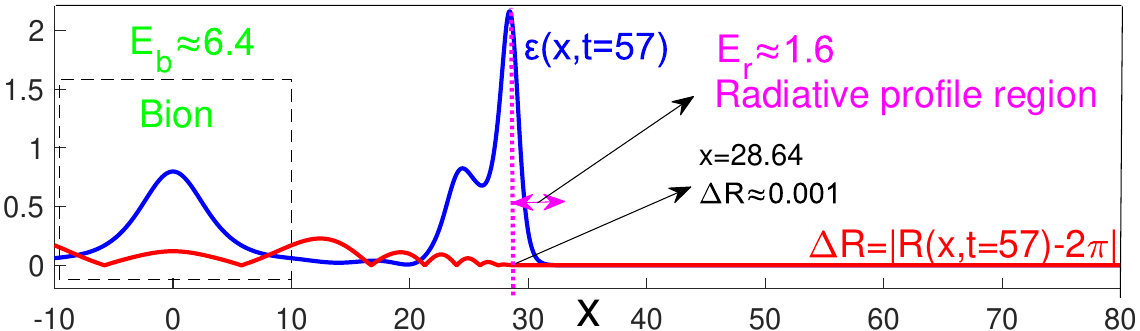}\\
			\includegraphics[width=120mm]{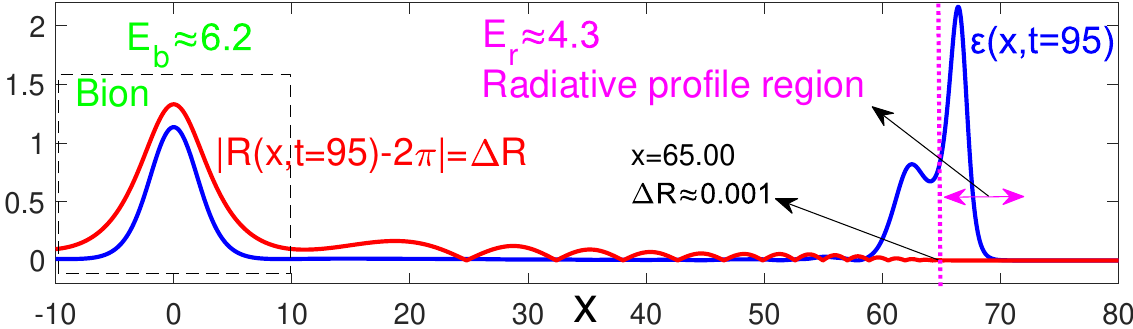}\\
			\includegraphics[width=120mm]{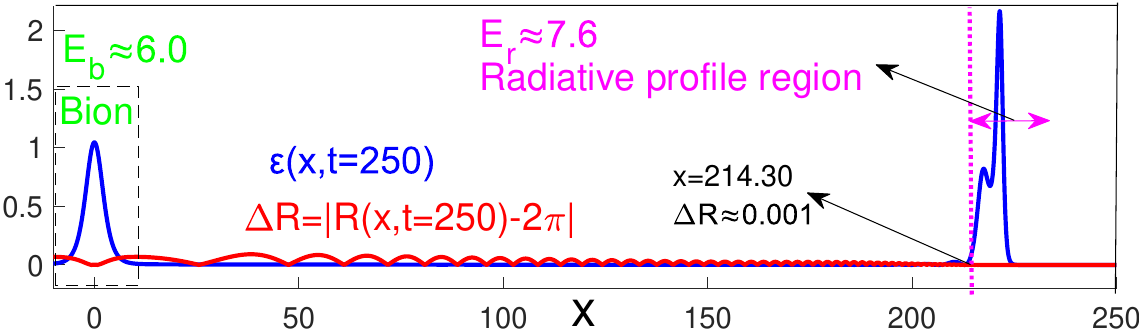}
		\end{tabular}
		\caption{Snapshots of energy density and field modulus for a collision at $\theta = 0.2\pi$ and $v = 0.7$. The radiative profile emitted to the right is highlighted at three time instances. The energy confined near the collision center (bion remnant) within the interval $[-10,10]$ is also shown.}
		\label{sk}
	\end{figure}

	To enable consistent comparisons, we select fixed reference times for all collisions. Specifically, the radiative profile energy is evaluated at \( t = (20/v) + 100 \) for Fig.~\ref{ft}(a) and at \( t = (20/v) + 220 \) for Fig.~\ref{ft}(b), corresponding to 100 and 220 time units after the collision moment, respectively.
	In practice, identifying a radiative profile involves detecting a region where \( R \) approaches \( 2\pi \). We therefore define this region by the condition \( |R - 2\pi| < 0.001 \), which indicates the presence of a radiative profile. 
	We verified numerically that moderate variations of this threshold (criterion) do not qualitatively modify the obtained results or the overall behavior of the radiative-profile energy curves.
	Fig.~\ref{sk} illustrates how this criterion is used to evaluate the radiative profile energy at three different time instances.

	Figures~\ref{ft} and \ref{sk} show that the radiative-profile energy evolves over time. The longer the post-collision evolution is followed, the more complete the radiative-energy measurement becomes. This time dependence weakens at late times, especially for low-velocity collisions. At high velocities, however, Fig.~\ref{ft} shows that the difference between measurements taken 100 and 220 time units after the collision can be significant. The corresponding high-velocity curves should therefore be interpreted with caution, particularly in regions exhibiting irregular features or downward trends. In principle, measurements at much later times, such as 1000 time units after the collision, would be more accurate, but they are restricted by numerical cost.

	Another key observation is that, following the collision and the formation of an unstable bion, additional low-amplitude pulses continue to emerge from the central oscillatory region. Numerically, these secondary structures initially appear ahead of the primary radiative profile before gradually synchronizing with the main radiative front. Consequently, energy accumulates behind the primary radiative profile over time, explaining the observed time dependence of the measured radiative-profile energy. In other words, the emitted radiation is continuously amplified by delayed energy transfer from the long-lived oscillatory structure formed at the collision center. This phenomenon is most clearly visible in Figs.~\ref{hjn}(b) and \ref{hjn}(c).

	One notable feature in Fig.~\ref{ft}(b) is the emergence of red critical velocities, such as a distinct jump at $v = 0.372$ for $\theta = 3\pi/12$, marking a transition from scattering to capture. The evaluation time of the radiative-profile energy influences the precise location of such transitions; longer evolutions yield more accurate identification of the critical velocities.
	
	\begin{figure}[htp]
		\centering
		\begin{tabular}{cc}
			\includegraphics[width=70mm]{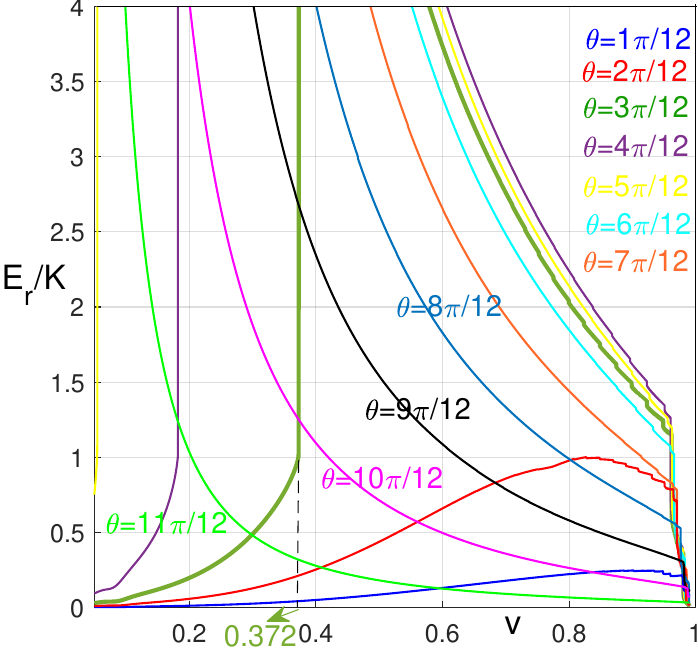} &
			\includegraphics[width=75mm]{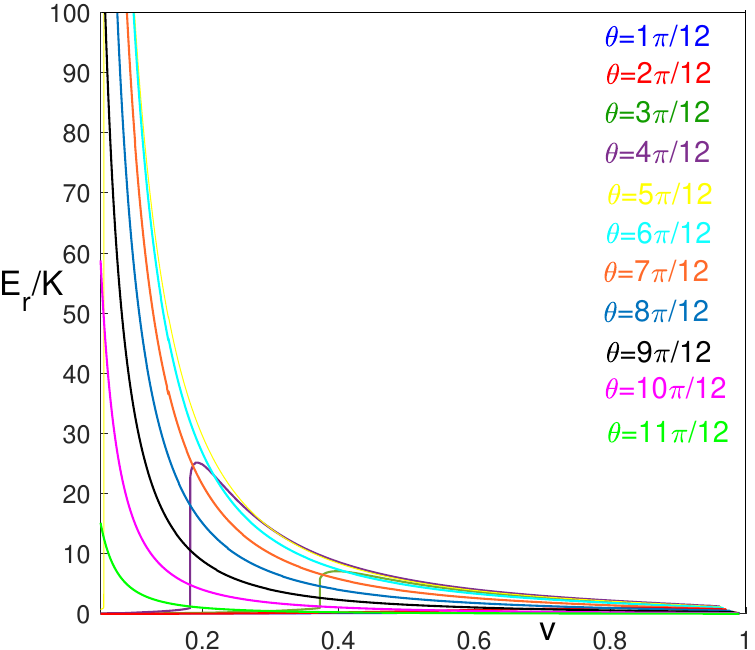}
		\end{tabular}
		\caption{Ratio of radiative profile energy to the initial kinetic energy as a function of velocity for various phase configurations.}
		\label{nnn}
	\end{figure}
	
	Finally, we define the ratio between the emitted radiative profile energy and the initial kinetic energy, as shown in Fig.~\ref{nnn}. This ratio serves as a diagnostic tool for distinguishing between capture and scattering outcomes.
	A ratio exceeding one confirms that capture has occurred, indicating that not only kinetic energy but also a portion of the kink rest energy has been radiated, making kink regeneration impossible. Conversely, if this ratio is below one after a sufficiently long evaluation time, it indicates scattering. This behavior is evident in Fig.~\ref{nnn}, particularly near the red critical velocity \(v = 0.372\) for \(\theta = 3\pi/12\), where the ratio approaches one. The high-velocity regime in Figs.~\ref{ft} and \ref{nnn} is therefore subject to significant numerical uncertainty.

	\section{Extreme Values} \label{sec6}
	
	\begin{figure}[htp]	
		\centering
		\begin{tabular}{c}	
			\includegraphics[width=130mm]{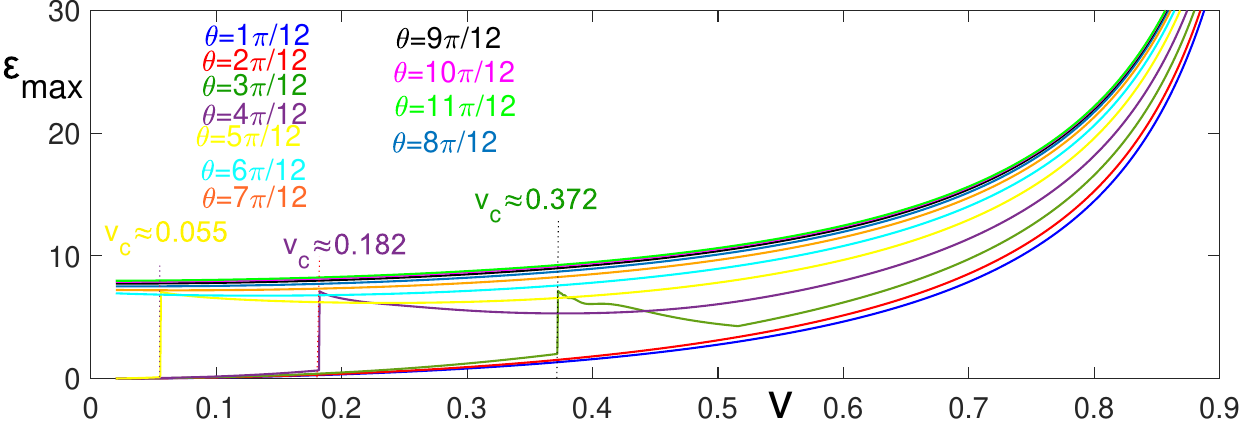}\\
			\includegraphics[width=130mm]{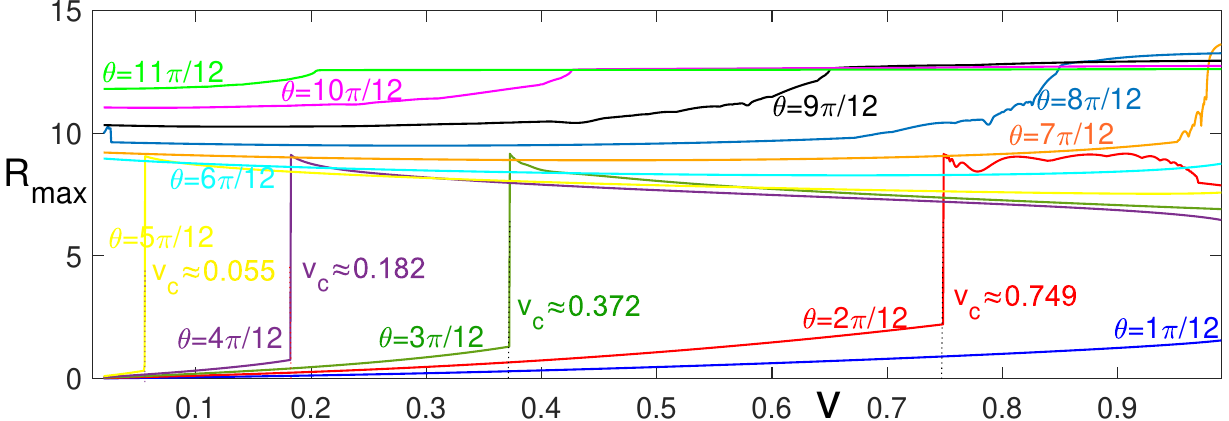}\\
			\includegraphics[width=130mm]{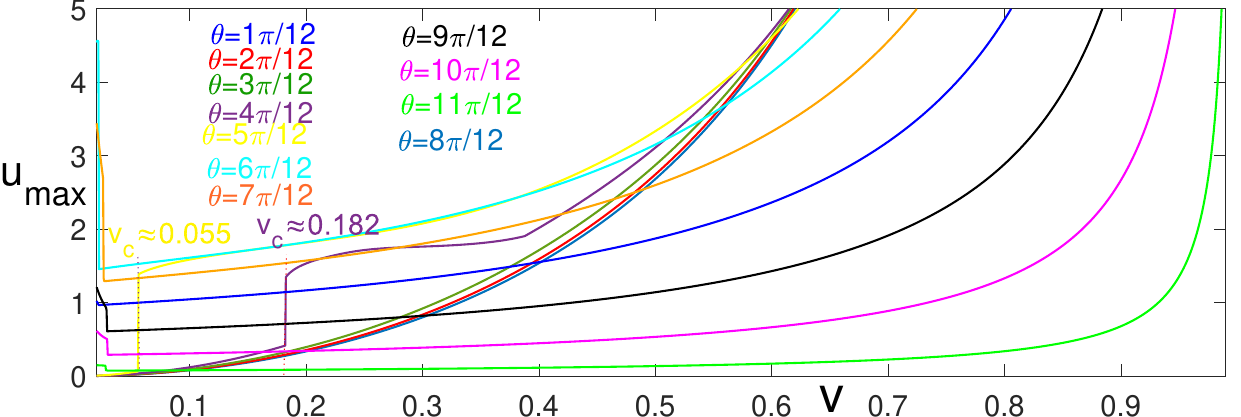}\\
			\includegraphics[width=130mm]{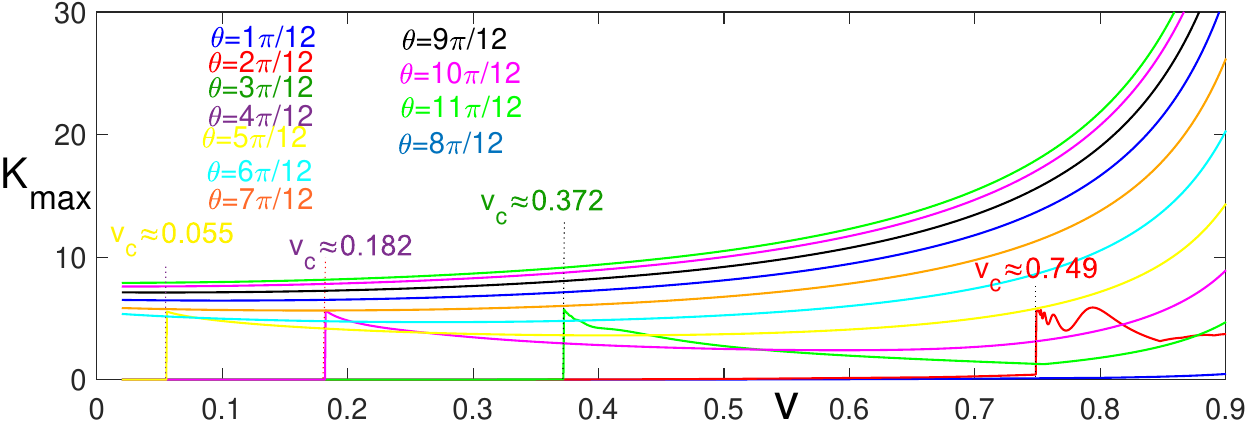}
		\end{tabular}
		\caption{Extreme values of various quantities at the center-of-mass point (\( x = 0 \)) plotted against velocity for different phase configurations.}
		\label{fg}
	\end{figure}
	
\begin{figure}[htp]	
		\centering
		\begin{tabular}{c}	
			\includegraphics[width=130mm]{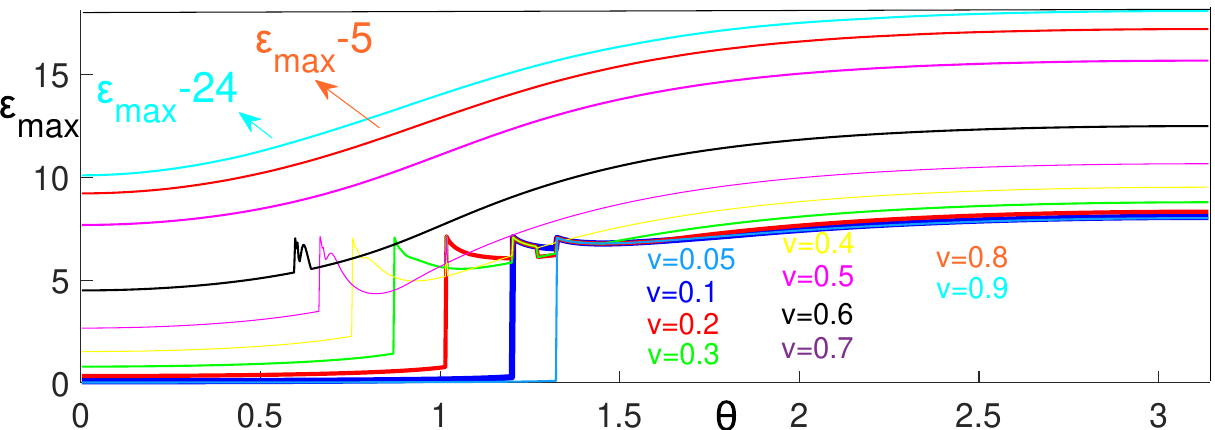}\\
			\includegraphics[width=130mm]{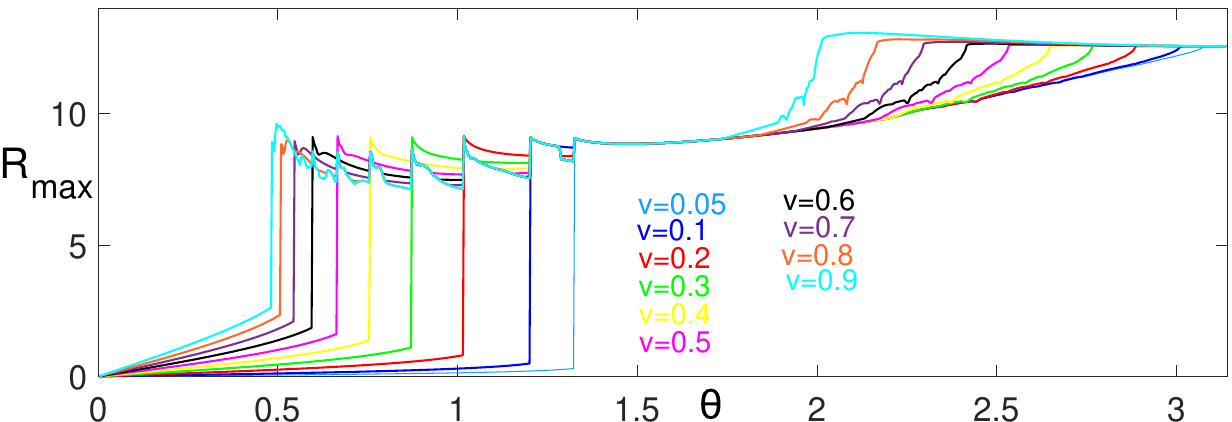}\\
			\includegraphics[width=130mm]{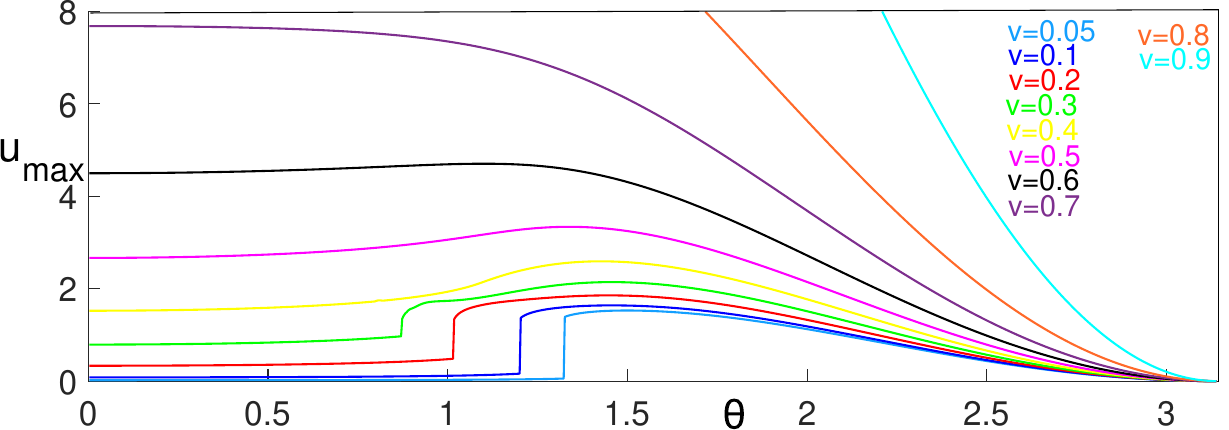}\\
			\includegraphics[width=130mm]{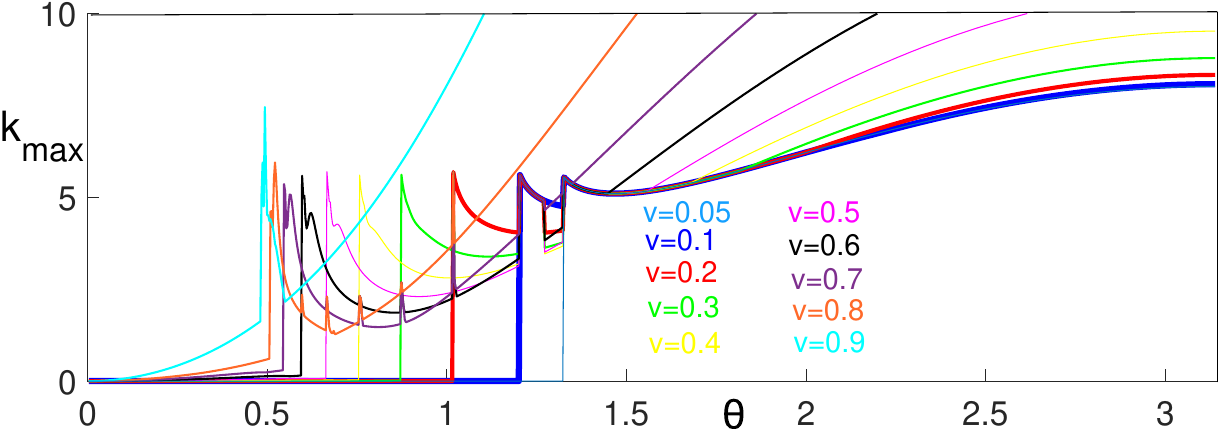}
		\end{tabular}
		\caption{Extreme values at the collision point plotted against phase for different fixed velocities.}
		\label{ggb}
	\end{figure}

	Determining the extreme (maximum or minimum) values of various quantities at the collision point (\( x = 0 \)) as functions of velocity and phase can significantly enhance our understanding of the interaction dynamics. Such analyses have been performed previously for systems with real kink solutions, including the sine-Gordon \cite{Saadatmand,Marjaneh}, double sine-Gordon (DSG) \cite{DSG6}, \(\varphi^4\) \cite{phi412,phi4153}, and \(\varphi^6\) models \cite{Moradi2}.
	In this study, we focus on the extrema of the modulus field (\( R \)), energy density (\( \varepsilon \)), and its kinetic and gradient components (\( k \) and \( u \)) at the collision point. Figure~\ref{fg} presents these extreme values as functions of velocity for eleven different phase values, while Fig.~\ref{ggb} shows their variation with phase for twelve fixed velocities.
	
	Examining the velocity-dependent plots of \( \varepsilon_{\text{max}} \), \( k_{\text{max}} \), \( u_{\text{max}} \), and \( R_{\text{max}} \) in Fig.~\ref{fg}, sharp discontinuities are observed at specific velocities, identified as red-type critical velocities. These jumps mark transitions from scattering to capture, highlighting the nonlinear sensitivity of the dynamics to both velocity and relative phase. Notably, \( k_{\text{max}}(v) \) and \( R_{\text{max}}(v) \) are particularly effective in detecting multiple critical velocities. Conversely, the \( u_{\text{max}}(v) \) curves are generally smoother, suggesting that the gradient part of the energy density is less sensitive to resonance and phase effects than the kinetic component.
	
	Turning to Fig.~\ref{ggb}, the behavior of these quantities as functions of the initial phase \(\theta\) for a range of velocities reveals additional complexity. At low phase values, the curves behave smoothly, increasing gradually with phase and indicating regular phase dependence. However, as the phase increases, the behavior becomes more irregular, with overlapping curves and multiple discontinuities that identify special phases associated with critical velocities.
	For instance, in the \( k_{\text{max}}(\theta) \) and \( R_{\text{max}}(\theta) \) plots, each corresponding to a fixed velocity, sharp jumps appear at particular phases. The phase at which such a discontinuity occurs defines the critical velocity for that phase, namely the velocity used in the plot.
	However, the exact type of critical velocity (red or blue) cannot be distinguished from these plots alone. Interestingly, jumps in the \( u_{\text{max}}(\theta) \) curves correspond exclusively to red-type critical velocities, emphasizing their importance in this phase-dependent context, contrary to their limited role in velocity-dependent plots.
	
	These results indicate that the collision dynamics are highly sensitive to the initial phase, sometimes even more so than to velocity. This is qualitatively different from the collision dynamics of the real SG model, where phase is essentially meaningless due to topological simplicity. In the complex sine-Gordon (CSG) model, each phase corresponds to a unique configuration of real and imaginary sub-kinks, creating a distinct internal structure and collision pathway. The pronounced phase dependence of the extreme values underscores the inherently multidimensional nature of collision dynamics in the CSG system, requiring joint consideration of velocity and phase spaces.

	\section{Summary and Conclusions} \label{sec7}

	In this work, we investigated the collision dynamics of complex kink solutions in the complex sine-Gordon (CSG) system, emphasizing their dependence on relative phase and velocity. The analytical structure of the model allows several classes of solitary-wave configurations, including complex kinks, radiative profiles, and Q-balls. Among these, complex kinks possess an internal degree of freedom characterized by a phase, leading to collision outcomes that differ substantially from those of their real counterparts.
	
	Through numerical simulations based on the discretized field equations, we explored a wide range of initial conditions. The results demonstrate that kink--kink collisions in the CSG model are highly sensitive to the relative phase. While in-phase collisions (\(\theta = 0\) or \(\pi\)) resemble the corresponding real SG dynamics, out-of-phase collisions (\(\theta \neq 0\) and \(\theta \neq \pi\)) emit radiative profiles and can exhibit capture scenarios, including the formation of bions or breather-like states.
	
	A key finding is the identification of two kinds of critical velocity separating scattering and capture regimes. In the red branch, increasing the initial velocity drives the system from scattering to capture. In the blue branch, increasing the velocity restores scattering. This dual categorization departs from the usual interpretation of a single critical velocity in real scalar-field models and reflects the role of the internal phase in CSG dynamics.
	
	We also analyzed the energy content of the emitted radiative profiles and demonstrated its time dependence. A substantial portion of the collision energy can be released through radiation, and the radiated energy depends sensitively on both the phase difference and the initial collision velocity. In addition, secondary low-amplitude radiation emitted from long-lived bion structures was observed during intermediate stages of the evolution before gradually synchronizing with the main radiative front. This behavior suggests delayed energy transfer and complex post-collision internal dynamics within the oscillatory structure.

	Finally, we examined several extreme quantities at the collision center, including the maximal energy density, kinetic and gradient energy densities, and field modulus. These diagnostics reveal sharp transitions corresponding to critical velocities, particularly in the phase-dependent plots. Thus, in the CSG system, interactions are governed by two independent parameters, velocity and phase, producing a richer solitonic interaction landscape than in real scalar-field models.

	These observations highlight the importance of internal degrees of freedom in nonlinear soliton interactions. In the CSG system, complex kink solutions may possess identical rest energies and similar spatial energy distributions while still exhibiting substantially different collision outcomes due to their relative phase structure. From this perspective, the internal phase acts as an additional dynamical parameter.

	Several questions remain open. Collisions between radiative profiles and complex kinks, and the dependence of those outcomes on phase, deserve a separate study. The overlap of some extreme-value curves in selected phase regions may also indicate hidden structural correlations. The regularity of the red and blue critical-velocity branches suggests that an analytical or semi-analytical relation between \(v_c\) and the phase difference may be derivable, for example through collective-coordinate or variational methods. Due to numerical constraints, extremely low velocities were excluded from the present study, and a comprehensive mapping of the oscillation periods \(T_L\) and \(T_S\) over the full velocity--phase plane remains an open problem. Collisions of three or more complex kinks with different phases and velocities may also reveal more intricate scattering patterns or chaotic behavior.
	
	Finally, the methodology and insights developed in this study may be extended to other classical scalar-field theories with kink solutions, such as the \(\varphi^4\), \(\varphi^6\), and double sine-Gordon models, where additional internal degrees of freedom may enrich the collision dynamics and generate new resonance structures or phase-sensitive critical phenomena.

	\section*{Acknowledgement}
	
	The authors wish to thank the Persian Gulf University Research Council for support.


\begin{thebibliography}{99}

		\bibitem{cosmology1} T. Vachaspati, \textit{Kinks and Domain Walls: An Introduction to Classical and Quantum Solitons} (Cambridge University Press, Cambridge, 2006).

		\bibitem{cosmology2} A. Vilenkin and E. P. S. Shellard, \textit{Cosmic Strings and Other Topological Defects} (Cambridge University Press, Cambridge, 1994).

		\bibitem{Manton} N. Manton and P. Sutcliffe, \textit{Topological Solitons} (Cambridge University Press, Cambridge, 2004).

		\bibitem{condense2} T. Dauxois and M. Peyrard, \textit{Physics of Solitons} (Cambridge University Press, Cambridge, 2006).

		\bibitem{Braun} O. M. Braun and Y. S. Kivshar, Phys. Rep. \textbf{306}, 1 (1998).

		\bibitem{optic3} G. P. Agrawal, \textit{Nonlinear Fiber Optics} (Academic Press, San Diego, 1995).

		\bibitem{bio} L. V. Yakushevich, \textit{Nonlinear Physics of DNA} (Wiley, Weinheim, 2006).

		\bibitem{phi44} D. K. Campbell, J. F. Schonfeld, and C. A. Wingate, Physica D \textbf{9}, 1 (1983).

		\bibitem{phi46} P. Anninos, S. Oliveira, and R. A. Matzner, Phys. Rev. D \textbf{44}, 1147 (1991).

		\bibitem{phi47} R. H. Goodman and R. Haberman, SIAM J. Appl. Dyn. Syst. \textbf{4}, 1195 (2005).

		\bibitem{phi61} P. Dorey, K. Mersh, T. Romanczukiewicz, and Y. Shnir, Phys. Rev. Lett. \textbf{107}, 091602 (2011).

		\bibitem{DSG2} V. A. Gani and A. E. Kudryavtsev, Phys. Rev. E \textbf{60}, 3305 (1999).






		\bibitem{Abraham} E. R. C. Abraham and P. K. Townsend, Phys. Lett. B \textbf{291}, 85 (1992).

		\bibitem{Leese} R. A. Leese, Nucl. Phys. B \textbf{366}, 283 (1991).

		\bibitem{Eto} M. Eto, T. Fujimori, S. B. Gudnason, K. Konishi, T. Nagashima, M. Nitta, K. Ohashi, and W. Vinci, J. High Energy Phys. \textbf{06}, 004 (2009).

		\bibitem{CTOPO} M. Eto, Y. Hirono, M. Nitta, and S. Yasui, Prog. Theor. Exp. Phys. \textbf{2014}, 012D01 (2014).
		
	    \bibitem{Nitta2022}	M. Nitta,  Physical Review D, \textbf{105(10)}  (2022), 105006.
	    
	    \bibitem{tONG} D. Tong, TASI Lectures on Solitons [arXiv:hep-th/0509216].
	    
	    

		\bibitem{MR} M. Mohammadi and N. Riazi, Prog. Theor. Exp. Phys. \textbf{2014}, 023A03 (2014).

		\bibitem{Tsumagari} M. I. Tsumagari, E. J. Copeland, and P. M. Saffin, Phys. Rev. D \textbf{78}, 065021 (2008).

		\bibitem{Bowcock} P. Bowcock, D. Foster, and P. Sutcliffe, J. Phys. A: Math. Theor. \textbf{42}, 085403 (2009).

		\bibitem{MohammadiQ} M. Mohammadi, Phys. Scr. \textbf{95}, 045302 (2020).

		\bibitem{Saadatmand} D. Saadatmand, S. V. Dmitriev, and P. G. Kevrekidis, Phys. Rev. D \textbf{92}, 056005 (2015).

		\bibitem{Marjaneh} A. M. Marjaneh, A. Askari, D. Saadatmand, and S. V. Dmitriev, Eur. Phys. J. B \textbf{91}, 22 (2018).

		\bibitem{DSG6} V. A. Gani, A. M. Moradi Marjaneh, and D. Saadatmand, Eur. Phys. J. C \textbf{79}, 620 (2019).

		\bibitem{phi412} A. Moradi Marjaneh, D. Saadatmand, K. Zhou, S. V. Dmitriev, and M. E. Zomorrodian, Commun. Nonlinear Sci. Numer. Simul. \textbf{49}, 30 (2017).

		\bibitem{phi4153} M. Mohammadi and E. Momeni, Chaos Solitons Fractals \textbf{165}, 112834 (2022).

		\bibitem{Moradi2} A. Moradi Marjaneh, V. A. Gani, D. Saadatmand, S. V. Dmitriev, and K. Javidan, J. High Energy Phys. \textbf{07}, 028 (2017).

	\end{thebibliography}
\end{document}